\documentclass{article}
\usepackage{graphicx}
\usepackage{amsmath}
\usepackage{amsfonts}
\usepackage{amssymb}

\begin{document}

\title{Some remarks about quantum diffusion for Hubbard models}
\author{R\'{e}my Mosseri\\Groupe de Physique des Solides, CNRS et Universit\'{e}s\\Paris 7 et 6, Tour 23, 2 place Jussieu \\75251 Paris Cedex 05, France\\e-mail: mosseri@gps.jussieu.fr}
\maketitle
\begin{abstract}
Some exact results are given, that connect, for a general magnitude of the
interaction term $U$, \ the repulsive and attractive Hubbard model, in terms
of the eigenspectra and quantum diffusion properties. In particular, it is
shown that, for some initial conditions, the quantum evolution cannot
differenciate between the attractive and repulsive models. These results apply
to both fermionic and bosonic models, in any dimension and for \ general
filling, as far as the underlying structure is bipartite.
\end{abstract}

\section{Introduction\bigskip}

Among the numerous models that have been studied for years to describe
correlated systems, the most popular is probably the Hubbard
model\cite{Hubbard}, parametrised by a first-neighbor hopping term $\beta$ and
on-site interaction energy $U$. The particles, electrons or bosons, can hop
between orbitals located at neighbouring sites of a $d$-dimensional lattice.
In one dimension, the exact eigenspectrum, for spin
${\frac12}$%
electrons, is known, thank to the well-known \ solution by Lieb and Wu
\cite{Lieb-Wu}. In higher dimension, no exact solution has been found, despite
the large efforts intended to fulfil this task\cite{efforts}. Therefore,
partial results are welcomed , as far as they can provide a better
understanding of this model. Usually, one considers separately the bosonic and
fermionic versions of this model, as well as positive (repulsive) or negative
(attractive) $U$ on-site interaction.

In this paper, I derive exact results that connect, for a general magnitude of
the interaction term $U$, \ the repulsive and attractive model, in terms of
their eigenspectra and quantum diffusion properties. It applies to both
fermionic and bosonic models, in any dimension and for \ general filling, as
far as the underlying structure is bipartite. There is already a long history
of works devoted to using all sorts of symmetries that simplify the problem,
some of which being intrinsic (like the SU(2) spin symmetry in the electron
hamiltonian), others being related to geometric properties of the underlying
lattice or graph (like the bipartiteness property), and/or to particular
electron filling(electron-hole transformation at half filling) \cite{Lieb
revue}\cite{Auerbach}. In section 2, a symmetry property of the spectrum is
derived, whch is certainly already known (see the side remark in ref
\cite{sutherland}), although I could not find it previously published in the
present form. To the best of my knowledge, the consequence on quantum
diffusion (section 3) is original. In any case, the demonstrations are not
technically difficult, but some results may appear, at first sight, rather
counterintuitive. Let me first summarize these results in two steps

(i) it will first be shown that on a bi-partite structure, for general filling
and dimension, the spectrum for positive $U$ is opposite to the spectrum for
negative $U$. In other words, to any eigenenergy $E$ for one model, it
corresponds (in a one-to-one manner) the opposite energy ($-E$) for the other
model. The relation between corresponding eigenstates is simple and will be explicited.

(ii) It is then possible to compare the time evolution, with the same initial
state, for the two models. It \ will be proved that, for some class of initial
conditions, these quantum evolutions cannot be distinguished.

It is this latter assertion which may appear at first sight as surprising. As
the simplest example, consider the case of two electrons of opposite spins
located on \ the same site of a one dimensional chain at $t=0$. If one
computes the wave packet at any positive time, one finds that the weight
(probability) of any ket in the tensor product basis is the same for $U<0$ and
$U>0$. The physical reason for that should appear clearly below.

\section{ A symmetry property for the eigenspectrum}

Let us first prove the first part about the eigenspectrum property. We shall
first construct a representation space $\mathcal{E}$ for the many-particle
Hilbert space, and map the many body problem onto a single-particle one in
$\mathcal{E}$.

\subsection{Warm-up: the two-electron case\bigskip}

To get familiar with this representation, let us describe the two-electron
case for a one-dimensional chain.With two electrons of opposite spin, we
consider one-particule states $\left|  l,\sigma\right\rangle $, where $l$ runs
on the sites and $\sigma=\pm1$ refers to the spin , which form a basis for the
one-particle Hilbert space, and construct 2-particle states in the tensor
product basis $\mathcal{B}$%

\[
\mathcal{B}=\left\{  \ \left|  M\right\rangle =\underset{}{\mid l,+\rangle
\otimes}\mid m,-\rangle=\mid l,m\rangle\text{, \quad}l,m\in\mathbf{Z}%
\right\}
\]

The representation space $\mathcal{E}$ can safely be taken as a 2 dimensional
square lattice, with a site ($l,m$) associated with the above ket $\mid
l,m\rangle$. Since each particle can hop only toward first neighbours in the
chain, the non vanishing hamiltonian elements are restricted to first
neighbours connections in $\mathcal{E}$, that is along the edges of the square
lattice. The standard Hubbard hamiltonian%

\begin{equation}
H=\beta\sum_{<lm>,\sigma}c_{l\sigma}^{+}c_{m\sigma}+U\sum_{i}c_{l+}^{+}%
c_{l+}c_{l-}^{+}c_{l-}=T(\beta)+V(U),
\end{equation}

maps onto a one-particle standard tight-binding on a square lattice, with
$\beta$ hopping terms, and a line of $U$ on-site potentials on the main diagonal,%

\begin{align}
H  &  =\beta\sum_{lm}(\left|  l,m\right\rangle \left\langle l+1,m\right|
+\left|  l,m\right\rangle \left\langle l-1,m\right|  +\left|  l,m\right\rangle
\left\langle l,m+1\right|  +\label{Hhop}\\
&  \left|  l,m\right\rangle \left\langle l,m-1\right|  )+U\sum_{l}\left|
l,l\right\rangle \left\langle l,l\right|  .
\end{align}

Note that, taking advantage of the periodicity along the diagonal (and
constructing Bloch sums labelled by $k$), we can map this problem onto a
family of one-dimensional problems with one impurity at the origin, and a
$k$-dependant hopping term. The standard Slater-Koster sum rule for this
impurity problem precisely corresponds to the Bethe-Ansatz condition for this
two-electron problem \cite{Caffa-Moss}.

Back to the 2 dimensional $\mathcal{E}$ space, we note the mirror symmetry
along the diagonal, which allows to split the solutions into fully symmetrical
and anti-symmetrical states, living repectively in $\mathcal{E}_{S}$ and
$\mathcal{E}_{A}$. $\mathcal{E}_{S}$ is a half space, containing the diagonal,
defined for instance as the points $\left\{  (l,m),l\geq m\right\}  $, a given
such point representing the symmetrized ket $\left(  \left|  l,m\right\rangle
+\left|  m,l\right\rangle \right)  /\sqrt{2}$. $\mathcal{E}_{A}$ is a also a
half space, excluding the diagonal, defined as the points $\left\{
(l,m),l>m\right\}  $ , a given point representing the antisymmetrized ket
$\left(  \left|  l,m\right\rangle -\left|  m,l\right\rangle \right)  /\sqrt
{2}$. It is then immediately clear that the antisymmetrical sector is
$U$-independent, as it is well known, and corresponds to the (addition)
spectrum (without double occupancy) found in the problem of two electrons of
equal spins. Indeed, in that case, Pauli principle imposes to restrict to the
space antisymmetrical sector $\mathcal{E}_{A}$.

The main points of interest here is that the two-particle problems has been
mapped onto a single particle problem in a larger space with a $U$ dependent
subspace, and that the bipartiteness property of the original structure (the
one-dimensional chain) is shared by the larger space (either $\mathcal{E},$
$\mathcal{E}_{A}$ and $\mathcal{E}_{S})$. These properties remain for the
general $N$-body problem, as we shall see now. Since spin is conserved in the
Hubbard hamiltonian, the $N$-particle problem can be separated into
independent problems where the number of particles of each spin is fixed. If
particles live in a d-dimensional space $\mathcal{F}$, the large space
$\mathcal{E}$ is just the euclidean sum of $N$ (a priori equivalent) copies of
$\mathcal{F}$. The spectral problem can be first studied (as will be done
below) in terms of a single particle problem in $\mathcal{E}$, selecting then
among the eigensolutions those satisfying the symmetries associated with
particle spins. An alternative way is to first reduce $\mathcal{E}$ by
symmetry, and then solve the reduced hamiltonian.

\subsection{Spectral symmetry}

Let us now derive our first result on the spectrum. We consider N particles,
of specified spin, in a hypercubic lattice $Z^{d\text{ }}$, and construct
$N$-particle states $\ \left|  M\right\rangle $, forming altogether a Hilbert
space basis $\mathcal{B}$ represented by a $Nd$ dimensional hypercubic lattice
$\mathcal{E}$ :%

\begin{equation}
\mathcal{B}=\left\{  \ \left|  M\right\rangle =\underset{i=1,N}{\bigotimes
}\mid r_{i},\sigma_{i}\rangle\right\}  , \label{tensor product basis}%
\end{equation}

More generally, if the one-particule states lives on a bi-partite graph
$\Lambda$ (not necessarily hypercubic), with edges connecting kets with non
vanishing hamiltonian matrix elements, so do the $N$-particule states. Indeed,
let us first define a sign function $s(r)$, taking the value $+1$ or $-1$
according to which sub-lattice of $\Lambda$,$\overrightarrow{r}$ belongs. In
the hypercubic lattice case with unit edge, $s(r)$ can for instance be defined
as
\begin{equation}
s(r)=(-1)^{\sum_{l=1,d}x_{l}} \label{signe one-electron}%
\end{equation}

where $x_{l\text{ }}$are the $d$ coordinates of $\overrightarrow{r}$. The sign
function $s(M)$ in the tensor product space is then simply defined as the
product of the one-particule sign functions $s(r)$ entering the tensor
product,
\begin{equation}
s(M)=\prod_{i=1,N}s(r_{i}) \label{signetensor product}%
\end{equation}

Now, looking for non-vanishing elements $\left\langle M^{\prime}\right|
H\left|  M\right\rangle $, with $M^{\prime}\neq M$, it is clear that they
occur only when one particule has jumped to a neighbouring site in $\Lambda$.
This implies $s(M^{\prime})=-s(M)$, which proves that the $N$-particule states
live on a bipartite structure. The interacting part translates into $N(N-1)/2$
hyperplanes of dimension $(Nd-1)$, with $U$ term as an on-site potential,
corresponding to kets in $\mathcal{B}$ such that two particles share the same
site in $\Lambda$. If these particles have equal spin, this manifold, that we
shall call a $U$-hyperplane, will be eliminated by a proper antisymetrisation
(which amount here to build new kets as the difference between kets that are
symmetrically related by this hyperplane mirror). Note that iteration of this
process vith the whole sets of hyperplanes for equal spins is nothing but
constructing Slater determinant states. The $U$-hyperplanes intersect along
hyperplanes of one dimension less, corresponding to states such that three
particles meet on a $\Lambda$ site, and so on, down to a single ''diagonal''
$U$-space of dimension $d$, corresponding to states such that all particles
share the same site. All these $U$-hyperplanes play a prominent role in the
bosonic case, while only the one with highest dimension occurs for the
electrons problem.

Since any symmetrization process done in $\mathcal{E}$ amounts to generate the
proper combination of sites related by the $U$-hyperplane mirrors, it is not
difficult to show that the $\mathcal{E}$ bi-partiteness property is shared by
any of the properly symmetrized or anti-symmetrized sets corresponding to the
particle statistics. Those willing to consider a specific type of particle
statistics, or to use more standard Fock spaces and second quantised operators
may \ prefer to redo the following computations in that framework, which will
eventually turn quicker. But I prefer here to analyze the spectral properties
in the (much) larger structure $\mathcal{E}$ , to stress that the properties
derived here ( on spectral symmetry and quantum diffusion) are statistics independent.

Let $\Delta(M)$ be the set of kets, different from $\left|  M\right\rangle ,$
and connected to \ it by $H$,
\begin{equation}
\Delta(M)=\left\{  \left|  M^{\prime}\right\rangle \in\mathcal{B}\text{,
}M^{\prime}\neq M\text{, and }\left\langle M^{\prime}\left|  H\right|
M\right\rangle \neq0\right\}  \text{,} \label{voisins}%
\end{equation}

and let $\left|  \Psi(E,+)\right\rangle $ be an eigenket of $H(+\left|
U\right|  )$, with eigenenergy $E$, whose decomposition reads
\begin{equation}
\mid\Psi(E,+)\rangle=\sum_{M\in\mathcal{B}}a_{M}^{E}\left|  M\right\rangle
\label{eigenket+}%
\end{equation}

By projecting the Schr\"{o}dinger equation onto the set of bra $\left\langle
M\right|  $, one gets a set of secular equations
\begin{equation}
\beta\sum_{M^{\prime}\in\Delta(M)}a_{M^{\prime}}^{E}=(E-f(M,U))a_{M}%
^{E}\text{, } \label{secular1}%
\end{equation}

where $f(M,U)=\left\langle M\left|  H\right|  M\right\rangle $ is related to
the number of multiply occupied 1-particle orbitals in $\left|  M\right\rangle
$. The only property which we now use is that $f$ is an odd function of $U$
(it is in fact a number time $U$). Let us now consider the ket
\begin{equation}
\left|  \Psi^{\prime}\right\rangle =\sum_{M\in\mathcal{B}}s(M)a_{M}^{E}\left|
M\right\rangle \label{psi_prime}%
\end{equation}

where $s(M)$ is the above defined sign function. Since $M$ and $\Delta(M)$
live on two different $\mathcal{E}$ subparts (for any $M$), $\mid\Psi^{\prime
}\rangle$ clearly satisfies the set of secular equations
\begin{equation}
\beta\sum_{M^{\prime}\in\Delta(M)}a_{M^{\prime}}^{E}=(-E-f(M,-U))a_{M}^{E}
\label{secular2}%
\end{equation}

which proves that $\left|  \Psi^{\prime}\right\rangle $ is an eigenstate of
$H(-\left|  U\right|  )$, with $-E$ as its eigenvalue: $\left|  \Psi^{\prime
}\right\rangle =\left|  \Psi(-E,-)\right\rangle $. This correspondance can be
carried out for all eigenvalues in the spectrum, and we can write
($\mathbf{Sp}$ denoting the set of eigenvalues in the spectrum)
\begin{equation}
\mathbf{Sp}(H,U)\equiv-\mathbf{Sp}(H,-U) \label{spectre1}%
\end{equation}

At this point, two remarks are worth pointing out

(i) If one consider only changing the hopping term sign, one easily finds (a
well known result) that
\begin{equation}
\mathbf{Sp}(H,-\beta,U)\equiv\mathbf{Sp}(H,\beta,U) \label{spectre2}%
\end{equation}

with, to a given eigenvalue $E$, $\left|  \Psi\right\rangle $ as an eigenket
for $H(\beta,U)$ and $\left|  \Psi^{\prime}\right\rangle $ for $H(-\beta,U)$

(ii) The hamiltonian $H$ is the sum of a kinetic part $T(\beta)$ and an
interacting part $V(U)$. In the $\mathcal{E}$ space, as said above, this
$V(U)$ part is an on-site interaction, like in impurity models, whose value
depends on the particular tensor product kets. The kinetic part takes a
constant value whenever two kets are connected by $H$. This reminds of the
one-electron so-called Anderson hamiltonian, but with two major difference:

- In the Anderson hamiltonian, the diagonal terms are randomly distributed,
while here the $U$ dependant terms are heavily correlated.

- One usually studies this Anderson hamiltonian onto lattices of physocal
interest (in dimension 1,2 or 3), why $\mathcal{E}$ is of high dimension.

It is nevertheless worth noticing that the above symmetry on the two spectra
apply as well to a standard Anderson hamiltonian, whenever all the random
diagonal terms have their sign changed simultaneously. At this point, one
should recall some recent work done on the two-electron problem (TIP) with
Hubbard hamiltonian, in the context of possible interaction induced
delocalization \cite{Shepelyanski}. If the (random, diagonal) potential
distribution law is symmetrical around zero, the properties of the model (
like the eigenstates localization length) should not change if all diagonal
terms have their sign changed. In such a case , one can predict that the
interaction induced delocalization around a given energy $E$ (with Hubbard
term $U$) should be identical to what occurs for $-E$ and $-U$

\section{Quantum Diffusion}

We now compare quantum diffusion (evolution under the time-dependant
Schr\"{o}dinger equation) for both sign of $U$ (attractive or repulsive
interaction), and with specified initial conditions. We first study the
simplest case, where $\left|  \Psi(t=0)\right\rangle =\left|  M\right\rangle
\in\mathcal{B}$. As usual, the time evolution is computed by expanding
$\left|  M\right\rangle $ in the complete set of eigenkets $\mathcal{E}_{\pm
}=\left\{  \left|  \Psi(E,\pm)\right\rangle \right\}  $. Let $W$ (inverse of
the above introduced $a_{M}^{E}$) be the unitary matrix transforming the
tensor product basis $\mathcal{B}$ into the eigenket basis $\mathcal{E}_{+}$.
Let us stress that $H$ being a real hamiltonian (which for instance would not
be the case in the presence of a magnetic field), it is always possible to
take $W$ as a real matrix, which is now assumed. Now, with $D=Card(\mathcal{B}%
)$, we can write
\begin{align}
\left|  \Psi(t=0)\right\rangle  &  =\left|  M\right\rangle =\sum
\limits_{l=1}^{D}W_{l}^{M}\left|  \Psi_{l}\right\rangle ,\text{ and}%
\nonumber\\
\left|  \Psi(t)\right\rangle  &  =\sum\limits_{l=1}^{D}W_{l}^{M}\exp
(-iE_{l}t/\hbar)\left|  \Psi_{l}\right\rangle \label{quantumdif1}%
\end{align}

where $l$ labels the set of eigenstates of $H(\left|  U\right|  )$. The
componant of $\left|  \Psi(t)\right\rangle $ on the ket $\left|  M^{\prime
}\right\rangle \in\mathcal{B}$ simply reads
\begin{subequations}
\begin{equation}
\langle M^{\prime}\left|  \Psi(t)\right\rangle =\sum\limits_{l=1}^{D}\left(
W_{l}^{M^{\prime}}\right)  ^{T}W_{l}^{M}\exp(-iE_{l}t/\hbar)
\label{quantumdif2}%
\end{equation}

Now, with the same initial condition, let us compute the corresponding ket at
time $t$ , $\left|  \Psi^{-}(t)\right\rangle $, for the hamiltonian
$H(-\left|  U\right|  )$. With the above discussed relations between the
eigenvalues and eigenkets, one easily obtains
\end{subequations}
\begin{equation}
\langle M^{\prime}\left|  \Psi^{-}(t)\right\rangle =s(M)s(M^{\prime}%
)\sum\limits_{l=1}^{D}\left(  W_{l}^{M^{\prime}}\right)  ^{T}W_{l}^{M}%
\exp(iE_{l}t/\hbar). \label{quantumdif3}%
\end{equation}

Note that the product $s(M)s(M^{\prime})$ takes only two values, $+1$ if $M$
and $M^{\prime}$ belong to the same sublattice, and $-1$ otherwise. So, in any
case, we find a simple correspondance between $\left|  \Psi(t)\right\rangle $
and $\left|  \Psi^{-}(-t)\right\rangle $,
\begin{equation}
\langle M^{\prime}\left|  \Psi(t)\right\rangle =\pm\langle M^{\prime}\left|
\Psi^{-}(-t)\right\rangle \label{quantumdif4}%
\end{equation}

Comparing both evolution for positive time, it comes
\begin{align}
\left|  \langle M^{\prime}\left|  \Psi^{-}(t)\right\rangle \right|  ^{2}  &
=\left(  \sum\limits_{l=1}^{D}\left(  W_{l}^{M^{\prime}}\right)  ^{T}W_{l}%
^{M}\exp(iE_{l}t/\hbar)\right)  \times\nonumber\\
&  \left(  \sum\limits_{l=1}^{D}\left(  W_{l}^{M^{\prime}}\right)  ^{T}%
W_{l}^{M}\exp(-iE_{l}t/\hbar)\right) \\
&  =\left|  \langle M^{\prime}\left|  \Psi(t)\right\rangle \right|  ^{2}
\label{quantumdif5}%
\end{align}

So quantum diffusion, at any time $t$, is identical (in terms of probabilities
with respect to any ket in the basis $\mathcal{B}$), for repulsive and
attractive interaction of the same magnitude ($\left|  U\right|  $). This
(surprising?) result is thus proved for any basis ket $\left|  M\right\rangle
,$ taken as an initial condition. It can be understood, on physical grounds,
by recalling first that quantum diffusion is done at constant energy. The
initial ket is delocalized in energy and its time evolution is influenced by
low and high energy eigenket contributions. So, states close to the ground
state energy for attractive $U$ correspond to more or less gathered particles,
while states with higher energy, eventhough we are in an attractive $U$ model,
correspond to more separated particles (this recalls the difference between
standard bonding and antibonding states for one-electron states). So high
energy states for attractive $U$ influences the quantum evolution in a similar
manner as low energy states for repulsive $U$, which qualitatively explain the
above result.

We now consider \ a more generic initial conditions for the quantum diffusion.
Let us split $\mathcal{E}$ (and therefore $\mathcal{B}$) into two parts
$\mathcal{E}_{A}$ and $\mathcal{E}_{B}$ , according to the bi-partite
decomposition, $\ ($corresponding to a basis partition $\mathcal{B}_{A}$ and
$\mathcal{B}_{B}$), and write
\begin{equation}
\left|  \Psi(t=0)\right\rangle =\sum_{M\in\mathcal{B}_{A}}b_{M,A}\left|
M\right\rangle +\sum_{M\in\mathcal{B}_{B}}b_{M,B}\left|  M\right\rangle .
\label{qd6}%
\end{equation}

It is then easy to write the components of $\left|  \Psi(t)\right\rangle $ in
the basis $\mathcal{B}$:
\begin{equation}
\langle M^{\prime}\left|  \Psi(t)\right\rangle =\sum\limits_{l=1}^{D}\left(
W_{l}^{M^{\prime}}\right)  ^{T}\exp(-iE_{l}t/\hbar)\left[  \sum_{M\in
\mathcal{B}_{A}}b_{M,A}W_{l}^{M}+\sum_{M\in\mathcal{B}_{B}}b_{M,B}W_{l}%
^{M}\right]  . \label{qd7}%
\end{equation}

The components of $\left|  \Psi^{-}(t)\right\rangle $ read accordingly
\begin{align}
\langle M^{\prime}\left|  \Psi^{-}(t)\right\rangle  &  =s(M^{\prime}%
)\sum\limits_{l=1}^{D}\left(  W_{l}^{M^{\prime}}\right)  ^{T}\exp
(iE_{l}t/\hbar)\times\label{qd8}\\
&  \left[  s(A)\sum_{M\in\mathcal{B}_{A}}b_{M,A}W_{l}^{M}+s(B)\sum
_{M\in\mathcal{B}_{B}}b_{M,B}W_{l}^{M}\right]
\end{align}

It is clear that now, $\left|  \langle M^{\prime}\left|  \Psi^{-}%
(t)\right\rangle \right|  ^{2}=\left|  \langle M^{\prime}\left|
\Psi(t)\right\rangle \right|  ^{2}$ only if the initial ket $\left|
\Psi(t=0)\right\rangle $ lives on one sub-structure only (either $A$ or $B$),
with all components $b_{M}$ in phase. This restricts the choice of initial
conditions for which this ($U,-U$) parallel behaviour occurs to a large but
definitively non generic set. Note however that one is often inclined, owing
to the linearity of Schr\"{o}dinger equation, to restrict the numerical
studies of quantum evolution to basis kets as initial conditions, which
precisely belong to this non generic category.

\bigskip

Acknowledgements: It is a pleasure to thank Claude Aslangul, Michel Caffarel,
Benoit Dou\c{c}ot and Julien Vidal for fruitful discussions.

\end{document}